\documentclass[sigconf]{acmart}

\usepackage{microtype}
\usepackage{subcaption}
\usepackage{tabulary}

\AtBeginDocument{%
  \providecommand\BibTeX{{%
    \normalfont B\kern-0.5em{\scshape i\kern-0.25em b}\kern-0.8em\TeX}}}

\copyrightyear{2020}
\acmYear{2020}
\acmConference[WWW '20 Companion]{Companion Proceedings of the Web Conference 2020}{April 20--24, 2020}{Taipei, Taiwan}
\acmBooktitle{Companion Proceedings of the Web Conference 2020 (WWW '20 Companion), April 20--24, 2020, Taipei, Taiwan}
\acmPrice{}
\acmDOI{10.1145/3366424.3384370}
\acmISBN{978-1-4503-7024-0/20/04}
\setcopyright{iw3c2w3}

\begin{document}

\title{Measuring Spatial Subdivisions in Urban Mobility \\ with Mobile Phone Data}

\author{Eduardo Graells-Garrido}
\orcid{0000-0003-0722-5969}
\affiliation{%
  \institution{Barcelona Supercomputing Center (BSC)}
  \city{Barcelona}
  \country{Spain}
}
\affiliation{
\institution{Universidad del Desarrollo}
\city{Santiago}
\country{Chile}
}
\email{eduardo.graells@bsc.es}

\author{Irene Meta}
\orcid{0000-0002-7220-119X}
\affiliation{%
  \institution{Barcelona Supercomputing Center (BSC)}
  \city{Barcelona}
  \country{Spain}
}
\affiliation{%
  \institution{Universitat Internacional de Catalunya (UIC)}
  \city{Barcelona}
  \country{Spain}
}
\email{irene.meta@bsc.es}

\author{Feliu Serra-Burriel}
\orcid{0000-0003-0879-8785}
\affiliation{%
  \institution{Barcelona Supercomputing Center (BSC)}
  \city{Barcelona}
  \country{Spain}
}
\affiliation{%
  \institution{Universitat Polit\`ecnica de Catalunya (UPC)}
  \city{Barcelona}
  \country{Spain}
}
\email{feliu.serra@bsc.es}

\author{Patricio Reyes}
\orcid{0000-0001-5364-9565}
\affiliation{%
  \institution{Barcelona Supercomputing Center (BSC)}
  \city{Barcelona}
  \country{Spain}
}
\email{patricio.reyes@bsc.es}

\author{Fernando M. Cucchietti}
\orcid{0000-0002-9027-1263}
\affiliation{%
  \institution{Barcelona Supercomputing Center (BSC)}
  \city{Barcelona}
  \country{Spain}
}
\email{fernando.cucchietti@bsc.es}

\renewcommand{\shortauthors}{E. Graells-Garrido et al.}

\begin{abstract}
Urban population grows constantly. By 2050 two thirds of the world population will reside in urban areas.
This growth is faster and more complex than the ability of cities to measure and plan for their sustainability. 
To understand what makes a city inclusive for all, we define a methodology to identify and characterize spatial subdivisions: areas with over- and under-representation of specific population groups, named \emph{hot} and \emph{cold} spots respectively. 
Using aggregated mobile phone data, we apply this methodology to the city of Barcelona to assess the mobility of three groups of people: women, elders, and tourists. 
We find that, within the three groups, cold spots have a lower diversity of amenities and services than hot spots. Also, cold spots of women and tourists tend to have lower population income.   
These insights apply to the floating population of Barcelona, thus augmenting the scope of how inclu\-si\-veness can be analyzed in the city.
\end{abstract}




\maketitle

\section{Introduction}
As of 2020, more than half of the population live in cities~\cite{UNpopulationdivision2018}, and by 2050, two thirds of the population will reside in urban areas~\cite{ritchie2018urbanization}.
In this scenario of urban growth, the United Nations have declared a goal for sustainable development that aims to make cities ``in\-clu\-sive, safe, resilient, and sustainable''~\cite{sdg}.
To reach these objectives, dis\-ci\-plines such as urbanism, architecture, ecology, sociology, and others provide frameworks to model the functioning of cities.
Typ\-i\-cal\-ly, the main data source for analysis is household and time-use surveys as well as travel diaries. Such instruments provide rich information that represents the general population of a city, which then informs urban design and policy making.

\sloppy However, the goals of improving inclusiveness and safety are limited by the purpose of traditional methods, because typical data sources tend to under-represent specific sub-populations, including women~\cite{chant2013cities} and elders~\cite{metz2000mobility}. For example, surveys fail to measure that women trips are shorter than those of men~\cite{blumen1994gender}, and how these trips are chained to others, partially due to household and care-taking purposes~\cite{mcguckin1999examining}. Likewise, elders also move in shorter trips, but, in contrast to younger people, their trip purposes are focused mostly on feeling independent and interacting with others in social situations~\cite{ziegler2011like}.
While it is known that traditional methods have these shortcomings, improving them to reach finer representation is expensive and impractical on a global scale.
Even though specific methods can be designed for under-represented groups, such efforts may miss the global picture, which includes the relationship be\-tween mobility of different po\-pu\-la\-tion groups. As a consequence, there is need of fine-grained city-scale data for the design of inclusive and safe urban spaces.
Finally, data biases often occur due to un\-der\-ly\-ing societal biases.
Biased data means incorrect population statistics which 
can mislead city planning and design into amplifying the problems they aim to fix~\cite{perez2019invisible}.

\sloppy Recent technological advances and the availability of non-traditio\-nal data sets have allowed to study urban phenomena at 
spatio-temporal granularities that traditional methods cannot. 
Mobile phone data, for example, allows a cost-effective way to perform studies about urban human mobility~\cite{calabrese2014urban}, as 
mobile operators already generate, store, and analyze the data for billing and marketing purposes.
The aggregation of digital traces from mobile phone usage was used to uncover data gaps in mobility~\cite{gauvin2019gender}, a
true seminal piece of work towards using this data source for inclusive cities.
Inspired by this line of work, we extend the scope to understand mobility aspects of three groups of urban visitors: women, elders, and tourists. 
Under the assumption that all people access the city equally, our research questions are: 
\emph{How to identify places with more (or less) presence of these groups than expected?} If these places can be pointed out, 
\emph{what characterizes them?} 
Our proposed method is a pipeline that starts on the definition of visitor metrics related to these three groups; 
then, we perform spatial analysis on these metrics to identify whether there is spatial concentration of visitors (or the lack thereof). If so, we identify areas with over-represen\-ta\-tion of these groups, or \emph{hot spots}, as well as the opposite, places with under-representation, or \emph{cold spots}.
We then proceed to use up-to-date information about income, amenities, and services in the city to characterize these areas based on their economic development and urban environment.

\begin{figure*}[t]
    \centering
    \includegraphics[width=\linewidth]{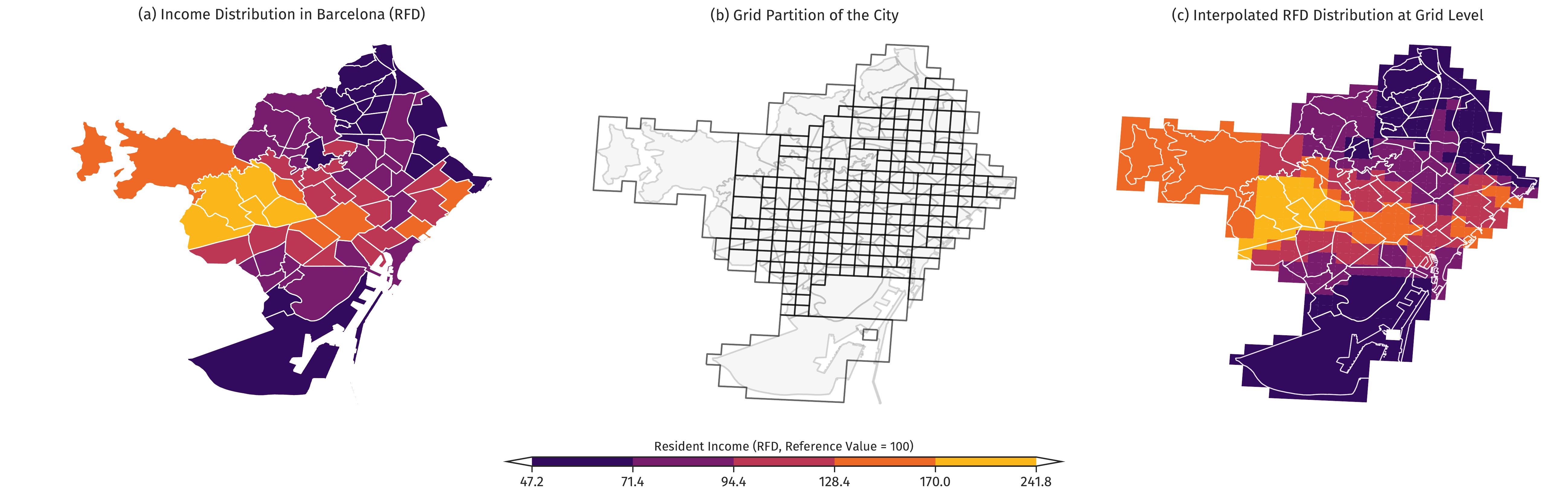}
    \caption{a) Map of neighborhoods with income data; b) Overlap of the grid areas with the neighborhoods; c) Spatial join of neighborhood income data with the grid.}
    \label{fig:bcn_districts_grid}
\end{figure*}

As a case study, we analyzed the city of Barcelona, the second largest city in Spain and one of the largest in Europe. The city is known for its urban planning tradition and was often ahead of its time compared to the Spanish general developments~\cite{marshall2004transforming}. However, there are still challenges in improving the city for everyone from a sustainable perspective:  overtourism~\cite{milano2019overtourism}, gender accessi\-bi\-li\-ty problems~\cite{i2016plan},  and one fifth of its inhabitants are elderly people~\cite{bcn2019salut}. 
In an effort to augment its understanding of the city's urban dynamics, the local government (\emph{Ajuntament de Barcelona}) acquired anonymi\-zed and aggregated mobile phone data from the mobile phone operator Vodafone. Access to this type of data  for planning is a clear opportunity to compare and advance over other data sources focused only on census and residential populations, as it could help to understand the mobility patterns of residents and non-residents alike.%, including tourists and commuters.

We find that the studied population subgroups behave differently. Cold spots for all groups are characterized by lower population income than hot spots, as well as less diversity of amenities and services. 
Hot spots for all groups are characterized for being less associated with public transport than the rest of the city. Urban infrastructure such as highways and the main streets of the city play a role when interpreting the locations of these areas of over- and under-representation, as cold spots tend to be outside of the area delimited by highways around the city whereas hot spots tend to be close to relevant primary streets.
As such, our work contributes:
(i) a methodology to identify and characterize hot and cold spots of the floating population in a city;
(ii) a case study applying the methodology to Barcelona.

We conclude the paper with a discussion focused on the impli\-ca\-tions in public policy and the usage of non-traditional data to solve the complex problems that affect cities today.

\section{Related Work}

Mobile phone data usually refers to the set of billing records from mobile phone networks, known as Data Detail Records (XDR)~\cite{calabrese2014urban}. 
Other types of non-traditional data that has been used to understand mobility include micro-blogging platforms with geo-location~\cite{mcneill2017estimating} or inferred user attributes regarding mobility~\cite{vasquez2019characterizing}; check-ins from location-based services~\cite{noulas2012tale}; and photos from photo-sharing services~\cite{beiro2016predicting}.
In comparison to these data sets, XDR allows a fine-grained analysis, not only in spatio-temporal aspects to, for instance, observe changes in mobility in short periods of time~\cite{graells2017effect}, but also in demographic ones, such as measuring the social diversity of visitors in shopping malls~\cite{beiro2018shopping}.

In terms of scope, our work is similar to recent efforts to un\-cov\-er gender gaps in urban mobility~\cite{gauvin2019gender}. 
The differences in our approach are two-fold. On the one hand, we use a data set that is aggregated from XDR in spatial and 
temporal aspects. As such, it does not include individualized information, allowing us to perform analysis at the area-level 
but not on the individual level. On the other hand, our methods rely primarily on established spatial analysis~\cite{moran1948interpretation,anselin1995local}, which brings a different perspective to the distance-based approach employed before. Then, our work contributes a different approach to an already identified problem, with extended coverage in terms of population groups, adding elders and tourists, and focusing on a different city, Barcelona.

\section{Context and Data Sets}

More than 1.6 million people reside in Barcelona. Its 100~Km$^2$area is composed of 12 districts, split into 73 neighborhoods.
Natural boundaries delimit the city: the Besos river limits the city at the north-west, and the Llobregat river does so at the south-west side. The Metropolitan Area is much wider, and it is impossible to distinguish the limit between Barcelona and the surrounding municipalities. The city extends on a mild slope from the sea (south-east) up to the edge of the Collserola mountain chain (north-west). The Collserola and the Montjuic (south) have limited the city expansion because of their relatively hard accessibility, and now are important areas of leisure and biodiversity within the city~\cite{marshall2004transforming}. 

The social aspects of mobility that affect subpopulations of the city~\cite{bcn2019salut,i2016plan} along with rising overtourism~\cite{milano2019overtourism} and alarming pollution levels have urged urban planners to focus on sustainability~\cite{pmu2019}. In this context, we focus on one of the qualities of sustainability, inclusiveness.

\begin{figure}[t]
    \centering
    \includegraphics[width=\linewidth]{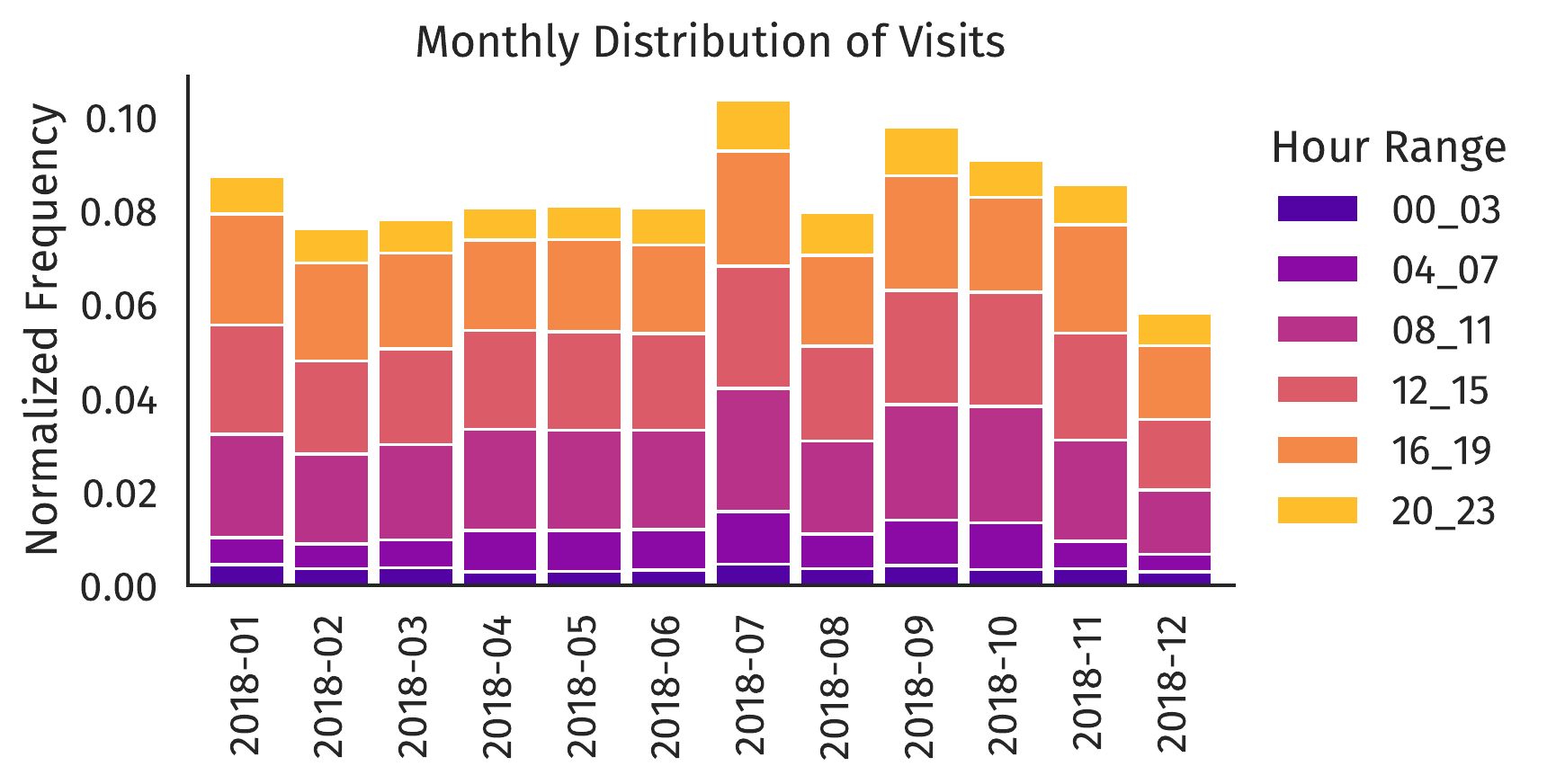}
    \caption{Normalized frequency of the sum of the monthly number of visitors captured in the dataset, provided by Vodafone to the Barcelona city hall. The colors stratify the sample into time-frames.}
    \label{fig:monthly_visits}
\end{figure}

\fussy

\paragraph{City Data}
The \emph{Ajuntament} provides open access to socio-de\-mo\-gra\-phic attributes at the neighborhood level at a yearly frequency (some metrics are quarterly), including income and house pricing among other things. All these variables are scaled to the mean, which allows us to compare the different neighborhood areas in relative terms. 
We measure income through the mean family income (RF, or \emph{Renda Familiar}),\footnote{\url{https://opendata-ajuntament.barcelona.cat/data/es/dataset/est-renda-familiar}} which contains mean income at the neigh\-bor\-hood level (see Figure~\ref{fig:bcn_districts_grid} (a) for its spatial distribution), normalized so that the whole city mean income equals 100. 

\paragraph{Mobile Phone Data}
The data obtained from the mobile phone operator consists of the number of visitors observed during the year 2018, at periods of four hours, grouped into 212 regions or cells (see Figure~\ref{fig:bcn_districts_grid} (b)). 
The number of visitors is defined as the total number of mobile phones active inside each region during each period. 
\emph{Active} means that the phone was initiated or received some activity (call, browse, text, etc) other than passive connections to the network.
This may introduce bias in the data, as  people do not call while driving or at night, or they connect to their home or job WiFi falsely indicating less presence. 
In addition to the total number of visitors, the operator also provides the number of visitors according to specific demographic characteristics, including gender (binary, female and male), age cohorts, and tourists (national and foreign). 
The determination of these characteristics and its ag\-gre\-ga\-tion into a number of visitors was made directly by the mobile operator
using activity criteria as well as billing and other in\-for\-ma\-tion when available.
\sloppy
In addition, cells with less than a given number of ob\-ser\-va\-tions during each period were discarded from the data and cells that consistently exhibited few observations (below 500 visitors) were consolidated into grouped regions. 

We estimated the total number of visits accounted per month and per hour range (see Figure~\ref{fig:monthly_visits}, normalized to avoid revealing commercially sensitive data). The number of total visitors per month lies within the same order of magnitude, with fluctuations that could be explained by changes in the market share of the operator and seasonal factors such as tourism in July. 

We estimated a mean income for each cell, defined as the weighted interpolation of the incomes in all areas that intersect with that cell (see Figure~\ref{fig:bcn_districts_grid} (c)).

\begin{figure}[t]
    \centering
    \includegraphics[width=\linewidth]{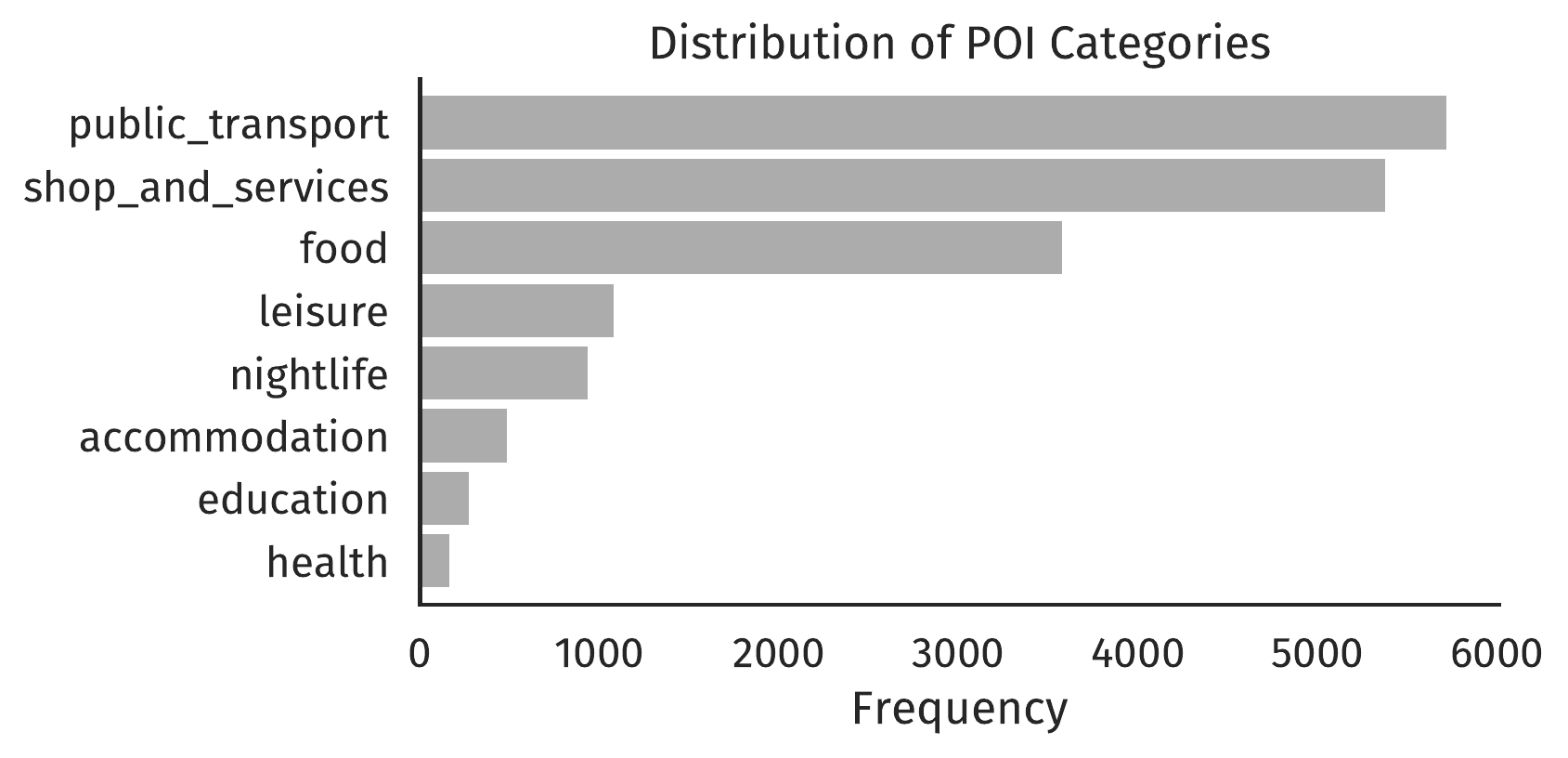}
    \caption{POI count per category. There are three main categories that represent more than 80\% of all the POI.}
    \label{fig:poi_count_per_category}
\end{figure}

\paragraph{Open Street Map}
To include aspects of the built environment and accessibility to amenities and services, we use data from Open\-Street\-Map (OSM).\footnote{http://openstreetmap.org} OSM provides spatial and geographical data contributed freely and voluntarily by its members, and it has been identified as an accurate source of urban information~\cite{haklay2010good}. 

From OSM we obtain Points of Interest (POIs) that allow us to understand part of the urban environment in our analysis. 
Points of interest (POIs) are geolocated attractors, such as shops, food spots, and tourist sights. We categorized most of POIs in the city as shown on Figure~\ref{fig:poi_count_per_category}. These include (sorted by descending frequency): \emph{public transport} (e.g., bus stops, metro stations), \emph{shops and services} (e.g., convenience shops, government offices, professional services), \emph{food} (e.g., caf\'es, restaurants), \emph{leisure} (e.g., natural attractions, parks, stadiums), \emph{nightlife} (e.g., bars), \emph{ac\-com\-mo\-da\-tion} (e.g., hotels), \emph{education} (e.g., universities, schools), and \emph{health} (e.g., hospitals). 

\null 

By analyzing the integration of these data sets, we aim to identify areas of the city with over- and under-representation of women, elders, and tourists, and characterize these areas according to their economic development and urban environment.

\section{Methods}

In this section we describe how we measure spatial subdivisions in urban mobility. 
First, we define the metrics we evaluate. These metrics are local, in the sense that they cover a specific point or area and do not consider the context or their surroundings.
Second, we perform a spatial analysis using established techniques, allowing us to take into account the spatial context and evaluate both local and global patterns to identify significant areas of over- and under-representation.
And third, we define how to characterize the significant areas with respect to economic development and urban environment.

\subsection{Cell Level Metrics}

In our context, the city is partitioned by a grid which comprises neigh\-bor\-ing cells. Cells may have edges and/or vertices in common but they do not overlap. Here we define cell-level metrics regarding the presence of women, elders, and tourists in them. 
The three metrics we define are: the women ratio $G$, the elder ratio $E$, and the tourist ratio $T$.

\paragraph{Women Ratio $G$}
This metric captures the ratio of female visitors in an area $i$ during the whole period under study. We first define the women ratio $G'$ as:
$$
G'_i = \frac{\text{\# women visiting area i}}{\text{\# total visitors in area i}}.
$$
Note that it is likely that the mobile operator has a non-representa\-tive sample of the population. For instance, the sample ratio at the city level may not be 1, even though the population ratio may be close to it. To counter this effect we define a standardized version of the women ratio as
$$
G_i = \frac{G'_i - \bar{G}'}{s(G')},
$$
where $s$ is the sample standard deviation function and $\bar{G}'$ corresponds to the mean of $G'_i$ for all the cells.
In this way, if $G_i = 0$, the area $i$ has a women ratio equivalent to the average of the city. 
Positive and negative values of $G$ indicate how many standard deviations the ratio deviates from the sample mean. 
Notice that, to focus on the floating population, we consider visitors between 8am and midnight.

\paragraph{Elder Ratio $E$}
In a similar way to $G'$, the elder ratio before standardization is defined as:
$$
E'_i = \frac{\text{\# elder visitors in area i}}{\text{\# total visitors in area i}}.
$$
We choose the threshold age to be considered elder as 65 years old or more, as defined by the \emph{Ajuntament}.\footnote{\url{https://ajuntament.barcelona.cat/personesgrans/es/canal/la-gent-gran-de-barcelona}}
Analogous to $G$, the metric $E$ is the standardized version of $E'$.

\paragraph{Tourist Ratio $T$}
Our last metric is similar to the previous ones, as it represents the proportion of tourists in an area:
$$
T'_i = \frac{\text{\# tourists (both foreign and national) in area i}}{\text{\# total visitors in area i}}.
$$
Analogous to $G$ and $E$, the metric $T$ is the standardized version of $T'$.

\subsection{Spatial Patterns}

Our aim is to find places where each population group of interest is over- or under-represented according to its floating population patterns, expressed in the metrics $G$, $E$ and $T$. To do so, we evaluate whether values of these metrics tend to concentrate in geographical terms, \emph{i.e.}, if nearby areas have similar values. The Moran's $I$ coef\-fi\-cient of spatial autocorrelation~\cite{moran1948interpretation} measures this concentration. It is defined as:
$$I = \frac{N}{W}\frac{\sum_i \sum_j w_{ij}(x_i - \bar{x})(x_j - \bar{x})}{\sum_i (x_i - \bar{x})^2},$$
where $N$ is the number of spatial units (in our case, grid cells) under analysis, $x_i$ is one of $\{G_i, E_i, T_i\}$, $w_{ij}$ encodes the spatial weight of cell $j$ into cell $i$, and $W$ is the sum of all spatial weights.
Note that $w_{ij}$ is a matrix where $w_{ii} = 0$. The value of $w_{ij}$ is a normalized version of the following schema:
\[
    w'_{ij}= 
\begin{cases}
    1 & \text{if area i and j are contiguous.}\\
    0               & \text{otherwise.}
\end{cases}
\]
Here, contiguity between cells is defined as sharing an edge or sharing a vertex. This is coherent when using grids composed of square cells, as it is possible to move from one square to another through a corner. 
Then, $w_{ij}$ is normalized in the following way:
$$
w_{ij} = \frac{w'_{ij}}{\sum_j w'_{ij}}. 
$$
With these definitions, $I = -1$ when the variable under analysis is perfectly dispersed in space, $I = 1$ when it is completely clustered, and $I = 0$ when values are randomly arranged.

Next, for each metric, if indeed there is spatial auto\-cor\-re\-la\-tion, we proceed to estimate Local Moran's $I$, a coef\-ficient that allows us to identify groups of areas that have high (or low) values that are surrounded by other areas with high (or low) values~\cite{anselin1995local}. It is defined as follows:
$$
I_i = \frac{x_i - \bar{x}}{s(x_i)^2} \sum^n_{j = 1, j \neq i} w_{ij}(x_j - \bar{x}),
$$
where $s(x_i)$ is the standard deviation of values of $x$ of  contiguous areas to area $i$.
Note that, in global and local $I$, significance is estimated through  permutation tests. Areas with significant high values of local $I$ are known as \emph{hot spots}, and areas with significant low values are known as \emph{cold spots}. The other areas  present neutral or average behavior.

\subsection{Characterizing Hot and Cold Spots}

After identifying areas of interest, our purpose is to characterize each type of spot. With this aim, we analyze the income of each spot and the availability of services and activities through Points of Interest (POIs).

\paragraph{Income}
We estimate a mean income for all spatial units. Under the null hypothesis that income is independent of spatial sub\-di\-vi\-sions, one would expect that population income in hot spots has the same distribution as population income in cold spots. We test this hypothesis by comparing the income in all hot spots with the income for all cold spots using the two-sample Kolmogorov-Smirnov (KS) test. This non-parametric test evaluates whether two underlying one-dimensional probability distributions that generated those samples differ. If the result of a test is significant, it means that there is evidence to reject the null hypothesis of same income for both types of area for a given visitor metric. 

\paragraph{Association and Diversity of Points of Interest}
Next, we estimate how each category of POIs is associated with each area.
This problem is analogous to document categorization in Information Retrieval where there are frequent words in many (if not all) doc\-u\-ments that do not necessarily characterize them. In our context, areas are analogous to documents, and POIs are analogous to words. For instance, bus stops may be available in all areas of the city, but some areas may have more bus stops than others, while having less POIs of other categories. In that case, these latter areas have a stronger association with bus stops than other areas.
Given that most areas have many kinds of POIs, we need to use a weighting schema that controls for frequency and variability. While a common technique to do so is Term Frequency--Inverse Document Frequency (TF-IDF), we resort to a technique that does not over-weight elements with low frequencies. This method is known as Log-Odds ratio with Uninformative Dirichlet Prior~\cite{monroe2008fightin}.  
It defines the weight of a word through the following point estimate:
\begin{equation*}
    \widehat{\delta}_{kw}^{(i)} = \log \begin{bmatrix}
\frac{(y_{kw}^{(i)} + \alpha_{kw}^{(i)})}
{(n_k^{(i)} + \alpha_{k0}^{(i)} - y_{kw}^{(i)}-\alpha_{kw}^{(i)})}
\end{bmatrix} - \log \begin{bmatrix}
\frac{(y_{kw} + \alpha_{kw})}
{(n_k + \alpha_{k0} - y_{kw}-\alpha_{kw})}
\end{bmatrix},
\end{equation*}
where $kw$ is the frequency of the POI type at cell $i$, and $\alpha$ is the prior distribution. 
Positive values of this metric indicate positive association, while negative values indicate disassociation. Thus, we would expect larger amounts of specific kinds of categories in specific regions and some other categories that are rare in other regions.  Values close to 0 indicate independence between the POI type and the cell. 

Ref.~\cite{gauvin2019gender} found relationships between accessibility gaps with the diversity of places, and between economic development and the diversity of social connections in places~\cite{eagle2010network}. To explore this potential relationship, for each cell $i$ we estimate its POI entropy, defined as the Shannon Information Entropy $H$:
$$
H_i = - \sum_c p_c \log p_c,
$$
where $c$ is a POI category, a $p_c$ is the fraction of POIs from category $c$ within cell $i$. 

\null 

By following this methodology, it is possible to identify spatial subdivisions in urban mobility according to who visits each area of the city, particularly women ($G$), elders ($E$), and tourists ($T$). The spatial subdivisions are defined as those areas identified as hot/cold spots of visits from these groups, which then can be characterized according to their economic development and urban environment.

\section{Results}

In this section we describe the results of applying our proposed methods to the data sets, with the aim of understanding spatial subdivisions in Barcelona, as seen from mobility data.

\begin{figure}[t]
    \centering
    \includegraphics[width=0.9\linewidth]{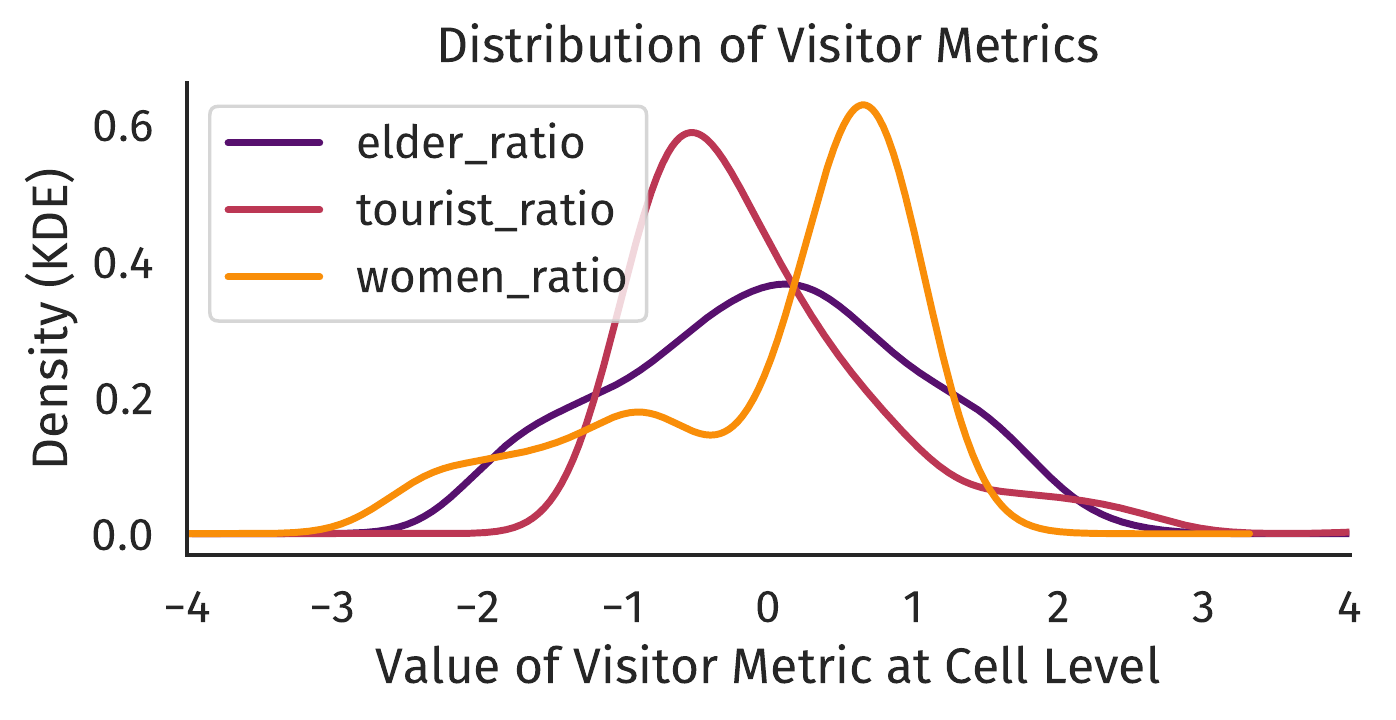}
    \caption{Probability density functions of cell-level metrics estimated with Kernel Density Estimation.}
    \label{fig:cell_level_metrics}
\end{figure}

\begin{figure*}[t]
    \centering
    \includegraphics[width=\linewidth]{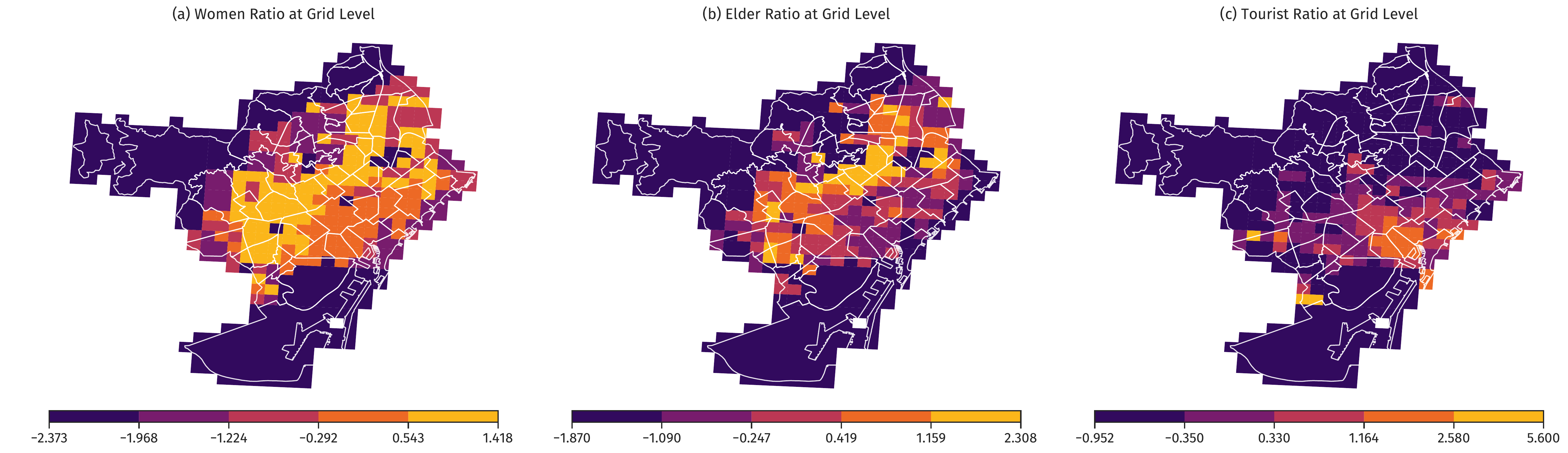}
    \caption{Maps of the spatial distribution for metrics $G$, $E$ and $T$. Color of cell $i$ corresponds to the value of $G_i$, $E_i$ and $T_i$ respectively. Notice that the scales are different for each one of the metrics.}
    \label{fig:metric_spatial_distribution}
\end{figure*}

\paragraph{Cell-level Metrics}
We estimated the women ratio $G$, elder ratio $E$, and tourist ratio $T$ for all cells in the grid of Barcelona. Of all cells, 195 have the same size. However, a few of them have bigger sizes, because they were merged by the mobile phone operator to ensure privacy. We considered the number of regular cells as a scaling factor (i.e., the most common) that would fit in a merged cell. Thus, we divided the value of each cell according to its scaling factor. 

The distributions of each observed metric (Figure~\ref{fig:cell_level_metrics}) have different shapes. The elder ratio distribution is unimodal with fat tails on both sides. The tourist ratio distribution is positively skewed, having a negative mode. Conversely, women ratio distribution is negatively skewed with positive mode. It has a group on the negative values but the majority of the values are positive.

\begin{figure*}
    \centering
    \includegraphics[width=\linewidth]{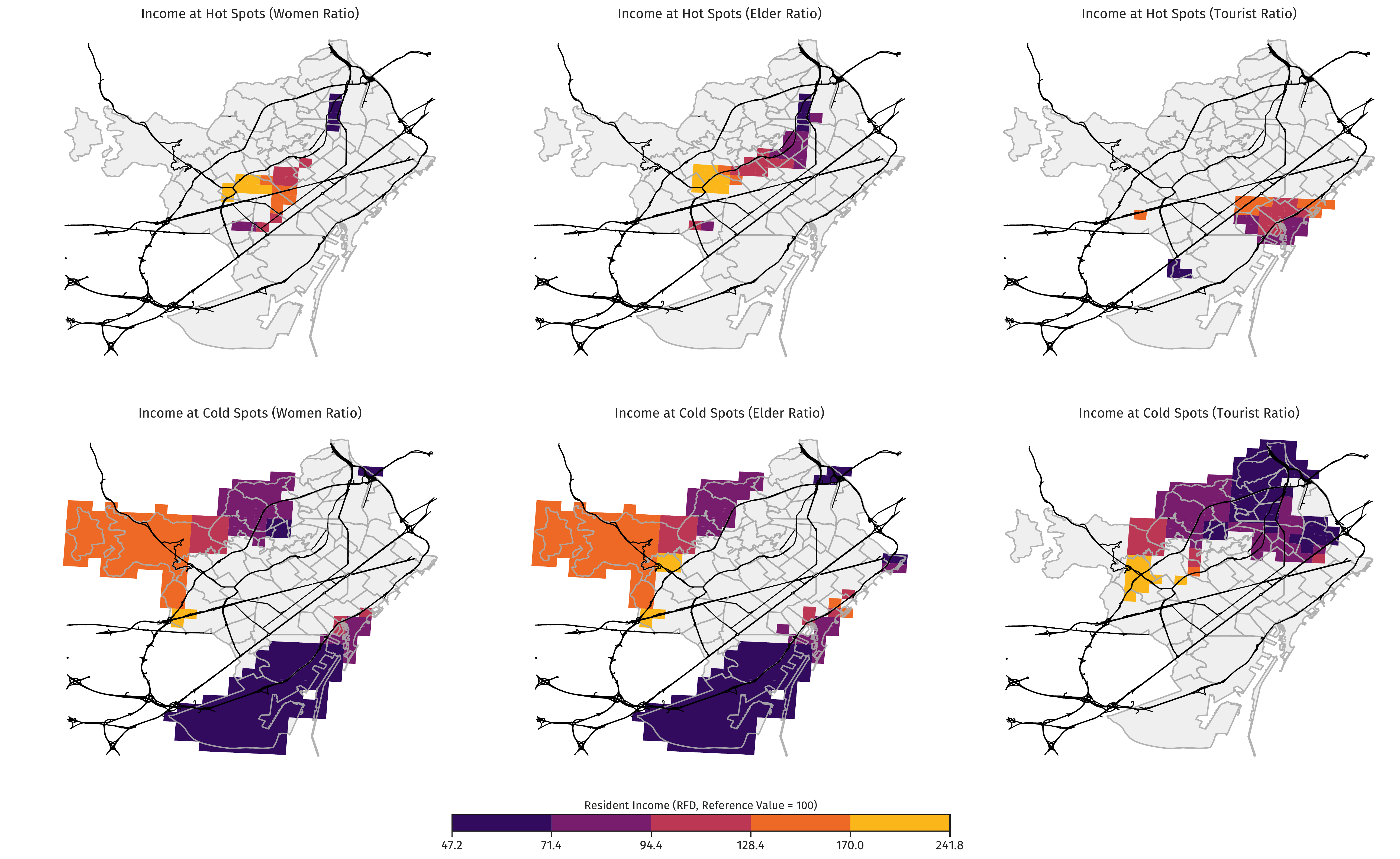}
    \caption{Hot/cold spots for each metric according to Local Mo\-ran's $I$ metric. 
    The first column is the gender ratio, the second column the elder ratio and the third column the tourist ratio. The top row contains hot spots, while the bottom row contains cold spots. The color represents the income of that area as in Figure \ref{fig:bcn_districts_grid}.}
    \label{fig:grid_local_moran}
\end{figure*}

\begin{figure*}
    \centering
    \includegraphics[width=0.95\linewidth]{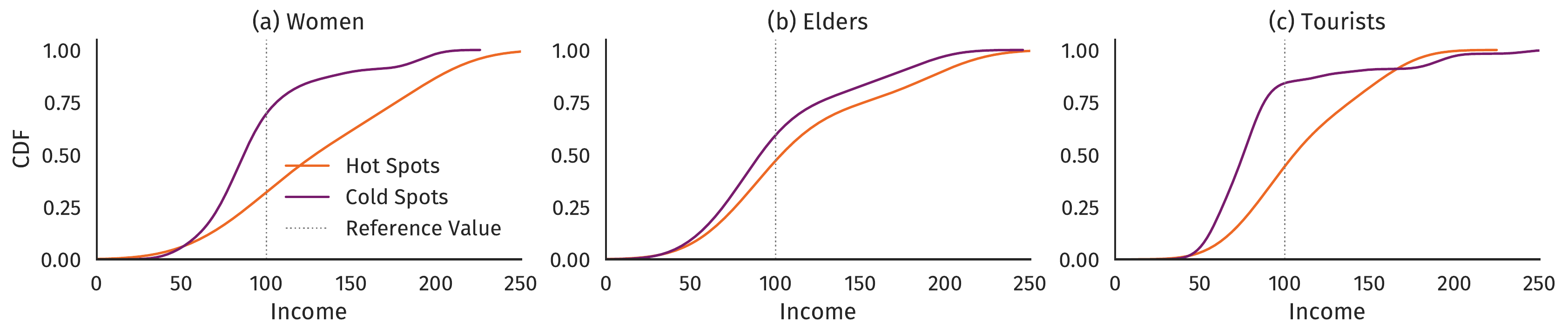}
    \caption{CDFs of the income on hot and cold spots for each of the metrics.}
    \label{fig:grid_spots_income}
\end{figure*}

\paragraph{Global Spatial Autocorrelation}
In spatial terms, the three metrics present spatial autocorrelation ($I_G = 0.25$, $I_E = 0.34$, $I_T = 0.39$, all significant with $p \leq 0.001$). 
Women and elders cover most of the densely populated areas of the city (see Figure~\ref{fig:metric_spatial_distribution} (a) and (b), respectively). However, the extent of elder concentration is smaller, thus having a greater autocorrelation than women. 
Tourists being the most concentrated group makes sense given the touristic at\-trac\-tive\-ness of the city, which tends to be concentrated in the historical districts, with few spots in other places such as the beaches, the highway that connects the airport to the city, and Barcelona's soccer team stadium (see Figure~\ref{fig:metric_spatial_distribution} (c)). 
Note, however, that all these concentrations are smaller than the concentration of income ($I_{RFD} = 0.83$, see Figure~\ref{fig:bcn_districts_grid}~(c)).

\paragraph{Local Spatial Autocorrelation}
Next, we estimated the Local Mo\-ran's~$I$ coefficient of each area of the city for $G$, $E$, and $T$. Figure~\ref{fig:grid_local_moran} shows the spatial location of all relevant areas, with hot spots of each metric in the top row, and cold spots of each metric in the bottom row. Color indicates the income level of each cell. 

Both $G$ and $E$ hot spots are located mostly above the Diagonal Avenue (the avenue the goes from west to east), and they overlap in three different sectors. The first point cor\-res\-ponds to the Sants area (west). The second sector is north-west of the city, right above the Diagonal Avenue. It cor\-res\-ponds to a middle and high income area of the city, while the third sector, to the north of the city, corresponds to a low income area. It is hard to interpret this over\-lap\-ping: we know the percentage of female population is larger with age~\cite{bcn2019salut}, but also that women are the ones who dedicate more time to care-taking activities~\cite{bcn2019salut}.
The $E$ hot spot shows a unique cluster, mainly just below the middle beltway. Within this hot spot, we observe heterogeneity on the socio-economic status, having the south-western part a much larger income ratio. 
The city center does not show under or over-representation of women and elders, as both hot and cold spots are absent there.
Conversely, the area in its totality is signalled as a hot spot for tourists ($T$), encompassing the Old district and part of the adjacent Eixample, below the Diagonal Avenue. This observation seems reasonable given the density of historical sites and leisure spots in the area. 
Two more isolated areas show a hot spot of tourist activity. 
The one on the west is nearby the Barcelona Football Club Stadium. Particularly, it is not the cell that contains the stadium, although it contains one of the metro stations that is closer to it. At the same time, it contains the southern campus of Universitat Politecnica de Catalunya, which may also receive foreigners regularly.
The other area is located on the south and contains the \emph{Fira Gran Via Barcelona}, the biggest venue of the city for international congresses and expositions (including the Mobile World Congress). This is expected given that the mobile operator may use roaming connections to identify foreign tourists.

The cold spots are mostly spread around the periphery of the city and have different levels of income.
There are three $G$ cold spots.
The smallest area is characterized by an infrastructural node that links motorways.
The largest area covers mixed income but also low population density, as it is located in the periphery of the city, near the mountains. 
The third area, on the south, covers a leisure sector (the Montjuic hill), a working sector (the Port), and a densely populated neighborhood, La Barceloneta, which holds many touristic attractions, including restaurants and the beach. 
The $E$ cold spots are similar to those of $G$. 
The western $T$ cold spot corresponds to the richest areas, and to the toll access tunnels from the valley behind the Collserola chain mountains. 
In the north side, $T$ cold spots comprehend the Sagrera high-speed train station and one of the main accesses to the city for road transport, with the infrastructural node between three motorways and two beltways. It is an area of low income and less leisure amenities than the rest of the city. Other areas of the city, such as the southern, are also characterized by similar income levels but they are not tourism cold spots.

\paragraph{Income Characterization}
There is a variety of income levels in hot and cold spots.
Women and tourist cold spots have different income distributions than their corresponding hot spots, according to a Kolmogorov-Smirnov (KS) test ($p_G = 0.004$, and $p_T < 0.001$, Bonferroni-corrected). Elders do not show that difference ($p = 0.798$, Bonferroni-corrected). To explore these differences visually, Figure \ref{fig:grid_spots_income} shows the cumulative distributions of income for hot/cold spots of each metric, estimated with Kernel Density Estimation (KDE). Hot spots tend to be shifted towards the larger income areas and cold spots appear to be on the low income areas.

\begin{figure*}
    \centering
    \includegraphics[width=\linewidth]{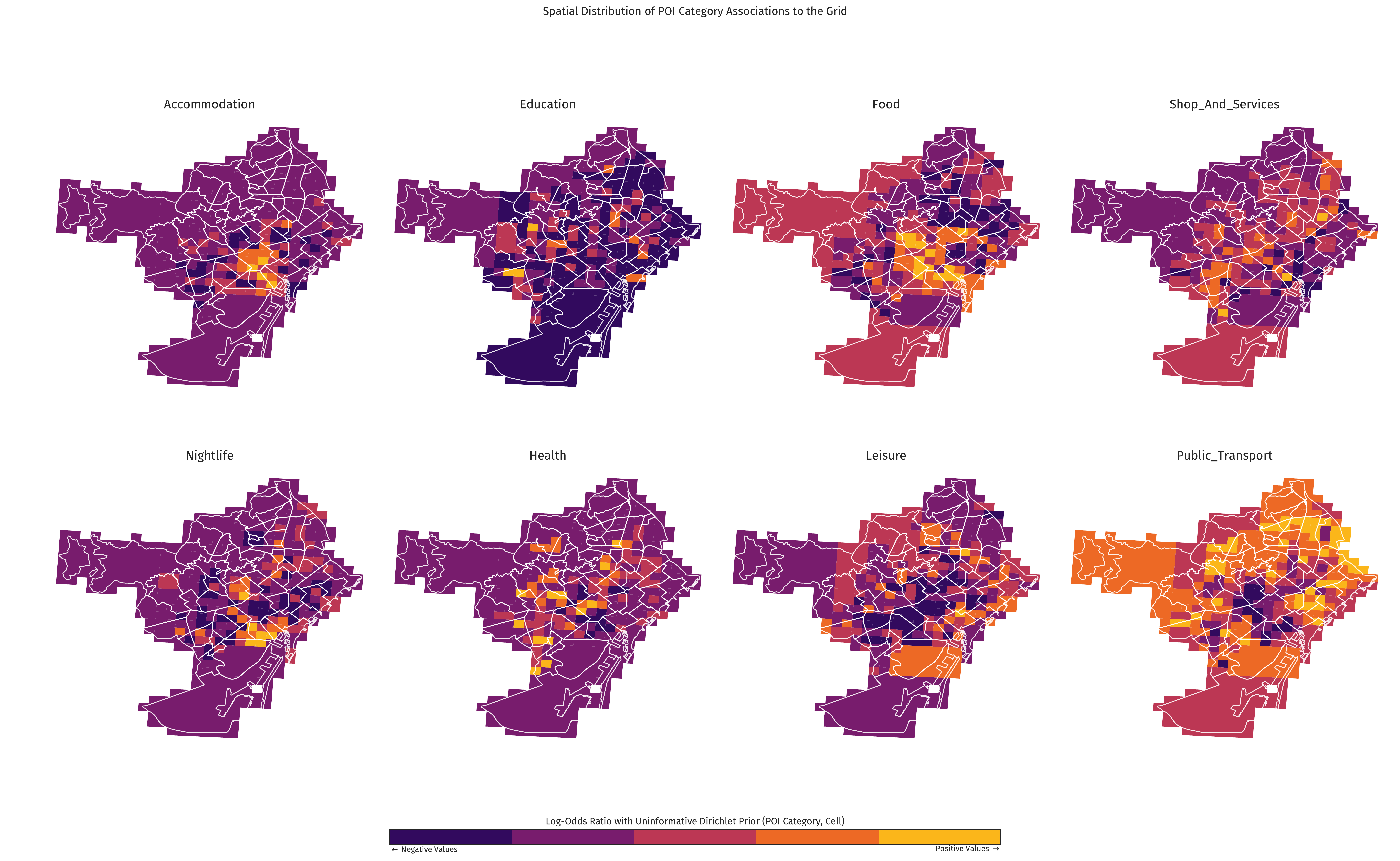}
    \caption{Log-odds Ratio with Uninformative Dirichlet Prior for each category of POI. Map colors according to the log-odds ratio value within each category.}
    \label{fig:poi_map_distribution}
\end{figure*}

\begin{figure*}
    \centering
    \includegraphics[width=0.9\linewidth]{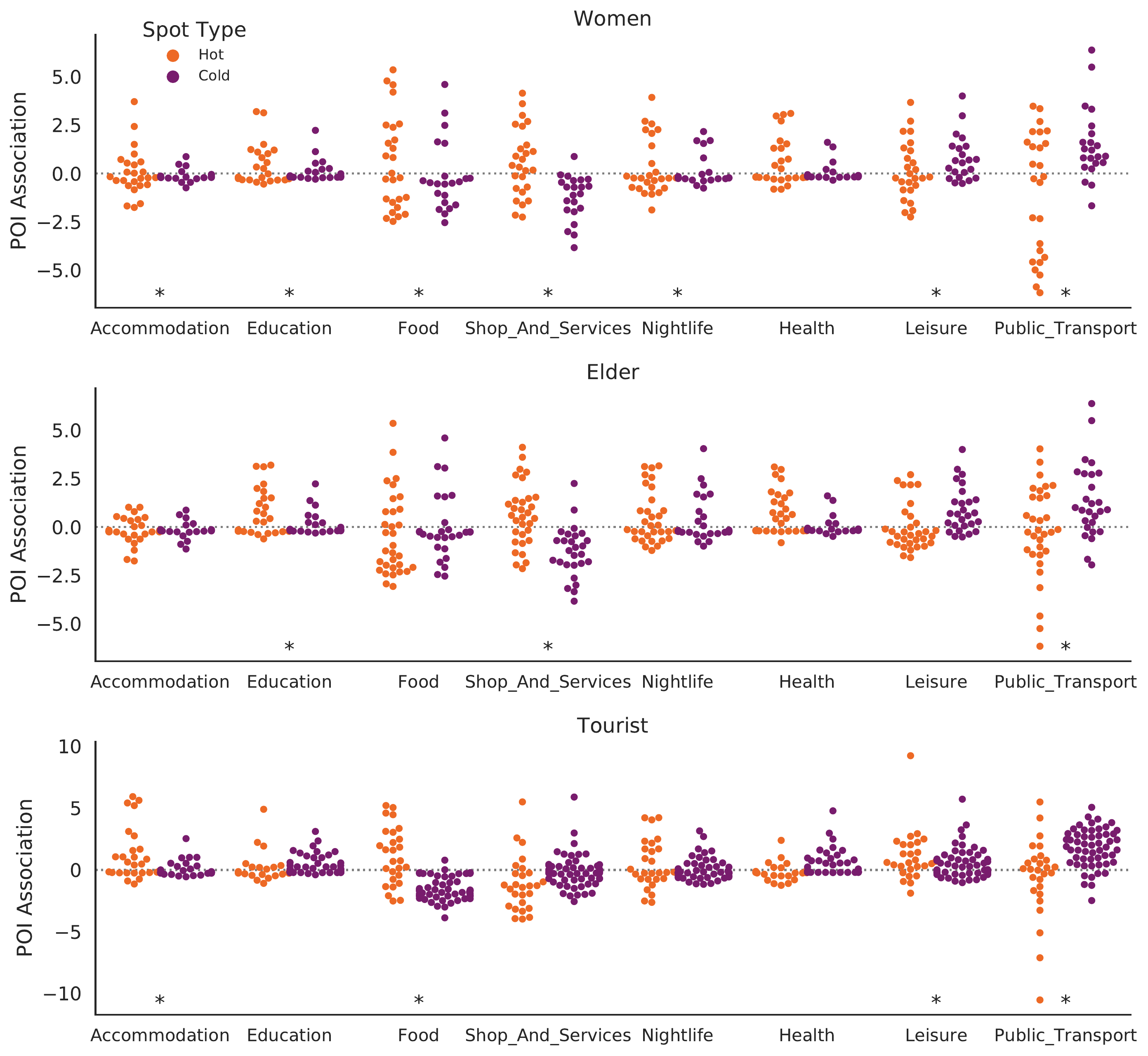}
    \caption{Swarm Plots of the POI association for each category of POI discerning hot and cold spots, for each kind of ratio. Plots marked with a star (*) indicate significant differences (according to K--S tests from Table~\ref{table:ks_tests_pois}) in POI association between Hot and Cold spots for the corresponding metric.}
    \label{fig:poi_association_boxplots}
\end{figure*}

\begin{table}[t]
\footnotesize
\begin{tabulary}{\linewidth}{p{0.5cm} LRRRR}
\toprule
Visitor Metric & POI Category & KS & \# Cells (Hot) &  \# Cells (Cold) &  $p$ \\
\midrule
   $G$ &  Accommodation  & 0.703 &  26 &  22 &        $< 0.001$ \\
   $G$ &  Education  & 0.633 &  26 &  22 &        0.001 \\
   $G$ &  Food  & 0.587 &  26 &  22 &        0.006 \\
   $G$ &  Shops~\&~Services    & 0.787 &  26 &  22 &        $< 0.001$ \\
   $G$ &  Nightlife & 0.549 &  26 &  22 &        0.019 \\
   $G$ &  Leisure  & 0.580 &  26 &  22 &        0.007 \\
   $G$ &  Public~Transport    & 0.657 &  26 &  22 &        0.001 \\
   $E$ &  Education & 0.471 &  32 &  27 &        0.043 \\
   $E$ &  Shops~\&~Services    & 0.678 &  32 &  27 &        $< 0.001$ \\
   $E$ &  Public~Transport    & 0.524 &  32 &  27 &        0.007 \\
   $T$ &  Accommodation  & 0.500 &  27 &  55 &        0.003 \\
   $T$ &  Food  & 0.707 &  27 &  55 &        $< 0.001$ \\
   $T$ &  Leisure  & 0.502 &  27 &  55 &        0.003 \\
   $T$ &  Public~Transport    & 0.593 &  27 &  55 &        $< 0.001$ \\
\bottomrule
\end{tabulary}
\caption{Kolmogorov-Smirnov tests (significance level \mbox{$p < 0.05$}, values show have been Bonferroni-corrected) for each of the metrics and significant POI categories on each studied subdivisions. Visitor metrics are women ratio ($G$), elder ratio ($E$), and tourist ratio ($T$).}
\label{table:ks_tests_pois}
\end{table}

\begin{figure*}
    \centering
    \includegraphics[width=0.95\linewidth]{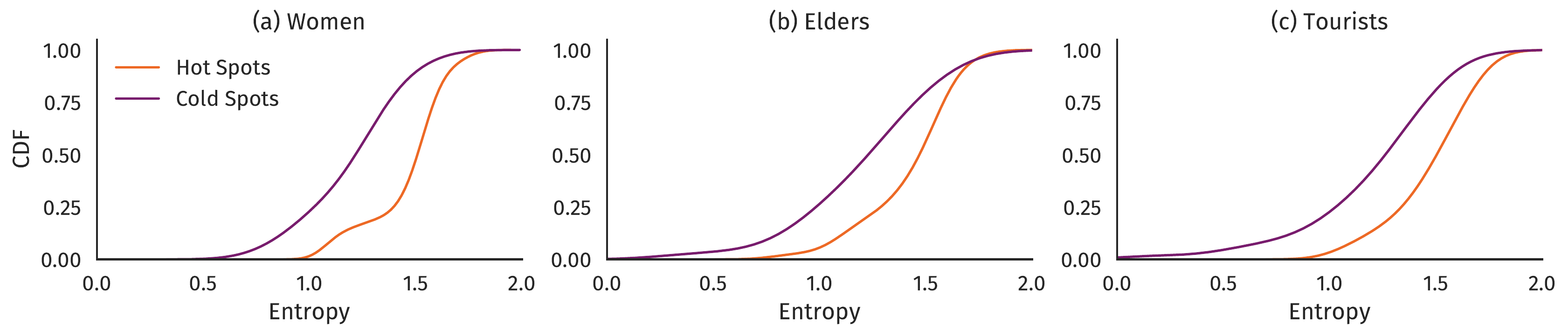}
    \caption{CDFs of the entropy for hot and cold spots on each of the ratios. 
    }
    \label{fig:entropies}
\end{figure*}

\paragraph{POI Characterization}
The distribution of POIs in the city exhibits different functional regions based on the activities and services available (see Figure~\ref{fig:poi_map_distribution}). 
Categories such as accommodation, food, and nightlife are more concentrated than the others, while health, shops and services and education are more scattered, indicating that most of the city has access to a diversity of amenities and services. 

To evaluate differences in POI 
(or \emph{amenities}) 
association between hot and cold spots, we performed pairwise KS tests for each POI category and each metric (see Table \ref{table:ks_tests_pois}). Then, we built swarm plots of each area type per metric, per POI category (see Figure~\ref{fig:poi_association_boxplots}). Every dot is a cell in a hot/cold spot of the associated variable, and its color represents its tendency (either hot or cold). Its $y$-position represents the corresponding POI association, while its $x$-position is only
for legibility.
Women ($G$) have the largest number of POI categories with significant differences between hot and cold spots association. Only the health category has the same distribution for hot and cold spots.
Elders ($E$) present differences in the distribution of POI association for education, shops and services, and public transport.
They present similar distributions to Tourists ($T$), where hot spots tend to be positively associated with amenities, except for the Public Transport category that presents some negative associations, similar to $G$ and $E$.
The hot spot association to amenities may be related to gender or age based mobility behaviours, 
regarding trip chaining and trip purposes; however, we lack a clear understanding of the disassociation to public transport, which is arguably un\-ex\-pected.
Tourists ($T$) present differences between hot and cold spots as\-so\-ci\-a\-tion on accommodation, food, leisure, and public transport. The first three categories
describe tourist attractors, as the hot spots are positively associated with these amenities. The public transport negative association to hot spots, similar to $G$ and $E$, may be ex\-plained due to the historic district being comprised mostly by pedestrian streets.
There are other associations that can be discussed, but we omit them due to space reasons.

Finally, regarding the diversity of POIs measured through entropy, cold spot areas are more associated with low diversity of POIs. The KS test was significant for the three pairwise comparisons ($p_G = 0.030$, $p_E = 0.005$, and $p_T < 0.001$, Bonferroni-corrected). The differences are illustrated through the KDE-based cumulative density functions on Figure~\ref{fig:entropies}. 

\null

In this section we explored how three population groups (women, elders, and tourists) were present in several areas of Barcelona in the year 2018. %Using aggregated mobile phone data, and using established spatial analysis methods, 
We observed that, indeed, there are areas of the city that tend to be visited by these groups (hot spots), as well as areas that tend to have an under-representation of them (cold spots), effectively creating subdivisions of over  and under-representation in the city. Income and the availability and diversity of POIs play a role in characterizing these relevant areas. The most salient characterizations are two. On the one hand, cold spots of activity for women and tourist visitors are associated with less population income. On the other hand, hot spots of the three types of visitors are associated with less public transportation. Cold spots of all types are associated with a lesser diversity of POIs. We discuss further implications of these results in the next section.

\section{Discussion and Conclusions}
\label{sec:discussion}
A city is experienced uniquely by each individual, although people with shared characteristics may experience it in similar ways.
Urban disciplines have been studying these experiences for decades with the goal of improving quality of life in cities through urban planning and design.
In this paper we have shown that aggregated mobile phone data allows us to identify relevant areas in terms of over- and under-representation of subpopulations such as women, elders, and tourists.
Being a cost-effective source of data, our proposal brings knowledge of which places are relevant in terms of presence (or absence) of people from these groups as well as what characterizes these places in terms of the urban environment. 
Then, our method\-ol\-ogy provides knowledge about under-represented groups in urban and policy design. 

We have shown that the places visited by specific groups are related to income and the presence and diversity of amenities and services. By using mobile phone data, we were able to present these insights for the floating population of Barcelona, in contrast to and thus complementing traditional data sources that focus on the resident population only.

Our work has two main limitations. 
First, the analysis is bound to the market share of the mobile phone operator, which is likely to be biased toward specific socio-economic and demographic groups. Given that the data is aggregated and anonymized, we cannot control for this fact. This motivated the usage of standardized metrics to entice a clearer interpretation of our results.
Second, there are intersections between the groups we analyzed, namely elderly female tourists. Hence, our analysis on income and POIs raises questions while providing preliminary answers which need further, deeper exploration, perhaps with more granular data.

In addition to improving the factors that limit the scope of this work, we devise three main lines of future work: 
the integration of additional area-level data sets, the definition of time-aware metrics, and multi-city analysis.
Including data about crime or health would improve the characterization of hot/cold spots. 

This aspect makes a time-aware analysis relevant, which would allow to measure the effect of urban interventions and seasonality according to our metrics.
Finally, the issues studied here are not exclusive to one city only. In order to advance on the path to inclusive, safe, resilient and sustainable cities, quantitative methods are required to compare cities within and between them, as well as fine-grained data sets to which apply these methods to. This would allow us to dis\-tin\-guish between systematic subdivisions and those specific to a city.

\begin{acks}
\sloppy
This project has received funding from the European Union's Hori\-zon 2020 research and innovation programme under grant agree\-ment No. 857191 (IoTwins project). E. Graells-Garrido was partially funded by CONICYT Fondecyt de Iniciaci\'on project \#11180913.
We ac\-knowl\-edge the following libraries used in the analysis: \textit{matplotlib}~\cite{hunter2007matplotlib}, \textit{seaborn}, \textit{PySAL}~\cite{rey2010pysal}, \textit{pandas}~\cite{mckinney2011pandas}, and  \textit{geopandas}.
Part of the map data used in this work is copyrighted by OSM contributors.
Thanks to Leo Ferres for insightful discussion, and to Xavier Paradis for his help in proofreading.
Finally, we thank the people from \emph{Ajuntament de Barcelo\-na} and Vodafone for providing access to the data and for useful dis\-cus\-sions.
\fussy
\end{acks}

\bibliographystyle{ACM-Reference-Format}
\bibliography{main}


\begin{thebibliography}{30}

%%% the Web address.

\ifx \showCODEN    \undefined \def \showCODEN     #1{\unskip}     \fi
\ifx \showDOI      \undefined \def \showDOI       #1{#1}\fi
\ifx \showISBNx    \undefined \def \showISBNx     #1{\unskip}     \fi
\ifx \showISBNxiii \undefined \def \showISBNxiii  #1{\unskip}     \fi
\ifx \showISSN     \undefined \def \showISSN      #1{\unskip}     \fi
\ifx \showLCCN     \undefined \def \showLCCN      #1{\unskip}     \fi
\ifx \shownote     \undefined \def \shownote      #1{#1}          \fi
\ifx \showarticletitle \undefined \def \showarticletitle #1{#1}   \fi
\ifx \showURL      \undefined \def \showURL       {\relax}        \fi
% invisible to TeX
\providecommand\bibfield[2]{#2}
\providecommand\bibinfo[2]{#2}
\providecommand\natexlab[1]{#1}
\providecommand\showeprint[2][]{arXiv:#2}

\bibitem[\protect\citeauthoryear{??}{sdg}{[n.d.]}]%
        {sdg}
 \bibinfo{year}{[n.d.]}\natexlab{}.
\newblock \bibinfo{title}{About the {S}ustainable {D}evelopment {G}oals}.
\newblock
  \bibinfo{howpublished}{\url{https://www.un.org/sustainabledevelopment/sustainable-development-goals/}}.
\newblock
\newblock
\shownote{Accessed: 2020-01-20.}


\bibitem[\protect\citeauthoryear{??}{pmu}{[n.d.]}]%
        {pmu2019}
 \bibinfo{year}{[n.d.]}\natexlab{}.
\newblock \bibinfo{title}{Urban Mobility Plan of Barcelona}.
\newblock
  \bibinfo{howpublished}{\url{https://www.barcelona.cat/mobilitat/es/actualidad-y-recursos/nuevo-plan-de-movilidad-urbana-2019-2024}}.
\newblock
\newblock
\shownote{Accessed: 2020-01-20.}


\bibitem[\protect\citeauthoryear{Anselin}{Anselin}{1995}]%
        {anselin1995local}
\bibfield{author}{\bibinfo{person}{Luc Anselin}.}
  \bibinfo{year}{1995}\natexlab{}.
\newblock \showarticletitle{Local Indicators of Spatial Association—LISA}.
\newblock \bibinfo{journal}{\emph{Geographical Analysis}} \bibinfo{volume}{27},
  \bibinfo{number}{2} (\bibinfo{year}{1995}), \bibinfo{pages}{93--115}.
\newblock


\bibitem[\protect\citeauthoryear{Beir{\'o}, Bravo, Caro, Cattuto, Ferres, and
  Graells-Garrido}{Beir{\'o} et~al\mbox{.}}{2018}]%
        {beiro2018shopping}
\bibfield{author}{\bibinfo{person}{Mariano~G Beir{\'o}},
  \bibinfo{person}{Loreto Bravo}, \bibinfo{person}{Diego Caro},
  \bibinfo{person}{Ciro Cattuto}, \bibinfo{person}{Leo Ferres}, {and}
  \bibinfo{person}{Eduardo Graells-Garrido}.} \bibinfo{year}{2018}\natexlab{}.
\newblock \showarticletitle{Shopping mall attraction and social mixing at a
  city scale}.
\newblock \bibinfo{journal}{\emph{EPJ Data Science}} \bibinfo{volume}{7},
  \bibinfo{number}{1} (\bibinfo{year}{2018}), \bibinfo{pages}{28}.
\newblock


\bibitem[\protect\citeauthoryear{Beir{\'o}, Panisson, Tizzoni, and
  Cattuto}{Beir{\'o} et~al\mbox{.}}{2016}]%
        {beiro2016predicting}
\bibfield{author}{\bibinfo{person}{Mariano~G Beir{\'o}},
  \bibinfo{person}{Andr{\'e} Panisson}, \bibinfo{person}{Michele Tizzoni},
  {and} \bibinfo{person}{Ciro Cattuto}.} \bibinfo{year}{2016}\natexlab{}.
\newblock \showarticletitle{Predicting human mobility through the assimilation
  of social media traces into mobility models}.
\newblock \bibinfo{journal}{\emph{EPJ Data Science}} \bibinfo{volume}{5},
  \bibinfo{number}{1} (\bibinfo{year}{2016}), \bibinfo{pages}{30}.
\newblock


\bibitem[\protect\citeauthoryear{Blumen}{Blumen}{1994}]%
        {blumen1994gender}
\bibfield{author}{\bibinfo{person}{Orna Blumen}.}
  \bibinfo{year}{1994}\natexlab{}.
\newblock \showarticletitle{Gender differences in the journey to work}.
\newblock \bibinfo{journal}{\emph{{U}rban {G}eography}} \bibinfo{volume}{15},
  \bibinfo{number}{3} (\bibinfo{year}{1994}), \bibinfo{pages}{223--245}.
\newblock


\bibitem[\protect\citeauthoryear{Calabrese, Ferrari, and Blondel}{Calabrese
  et~al\mbox{.}}{2014}]%
        {calabrese2014urban}
\bibfield{author}{\bibinfo{person}{Francesco Calabrese}, \bibinfo{person}{Laura
  Ferrari}, {and} \bibinfo{person}{Vincent~D Blondel}.}
  \bibinfo{year}{2014}\natexlab{}.
\newblock \showarticletitle{Urban Sensing Using Mobile Phone Network Data: A
  Survey of Research}.
\newblock \bibinfo{journal}{\emph{ACM Computing Surveys (CSUR)}}
  \bibinfo{volume}{47}, \bibinfo{number}{2} (\bibinfo{year}{2014}),
  \bibinfo{pages}{1--20}.
\newblock


\bibitem[\protect\citeauthoryear{Chant}{Chant}{2013}]%
        {chant2013cities}
\bibfield{author}{\bibinfo{person}{Sylvia Chant}.}
  \bibinfo{year}{2013}\natexlab{}.
\newblock \showarticletitle{Cities through a ``gender lens'': a golden ``urban
  age'' for women in the global South?}
\newblock \bibinfo{journal}{\emph{Environment and Urbanization}}
  \bibinfo{volume}{25}, \bibinfo{number}{1} (\bibinfo{year}{2013}),
  \bibinfo{pages}{9--29}.
\newblock


\bibitem[\protect\citeauthoryear{de~Barcelona}{de~Barcelona}{2019}]%
        {bcn2019salut}
\bibfield{author}{\bibinfo{person}{Consorci~Sanitari de Barcelona}.}
  \bibinfo{year}{2019}\natexlab{}.
\newblock \showarticletitle{La Salut a Barcelona 2018}.
\newblock  (\bibinfo{year}{2019}).
\newblock


\bibitem[\protect\citeauthoryear{Eagle, Macy, and Claxton}{Eagle
  et~al\mbox{.}}{2010}]%
        {eagle2010network}
\bibfield{author}{\bibinfo{person}{Nathan Eagle}, \bibinfo{person}{Michael
  Macy}, {and} \bibinfo{person}{Rob Claxton}.} \bibinfo{year}{2010}\natexlab{}.
\newblock \showarticletitle{Network Diversity and Economic Development}.
\newblock \bibinfo{journal}{\emph{Science}} \bibinfo{volume}{328},
  \bibinfo{number}{5981} (\bibinfo{year}{2010}), \bibinfo{pages}{1029--1031}.
\newblock


\bibitem[\protect\citeauthoryear{Gauvin, Tizzoni, Piaggesi, Young, Adler,
  Verhulst, Ferres, and Cattuto}{Gauvin et~al\mbox{.}}{2019}]%
        {gauvin2019gender}
\bibfield{author}{\bibinfo{person}{Laetitia Gauvin}, \bibinfo{person}{Michele
  Tizzoni}, \bibinfo{person}{Simone Piaggesi}, \bibinfo{person}{Andrew Young},
  \bibinfo{person}{Natalia Adler}, \bibinfo{person}{Stefaan Verhulst},
  \bibinfo{person}{Leo Ferres}, {and} \bibinfo{person}{Ciro Cattuto}.}
  \bibinfo{year}{2019}\natexlab{}.
\newblock \showarticletitle{Gender gaps in urban mobility}.
\newblock \bibinfo{journal}{\emph{arXiv preprint arXiv:1906.09092}}
  (\bibinfo{year}{2019}).
\newblock


\bibitem[\protect\citeauthoryear{Graells-Garrido, Ferres, Caro, and
  Bravo}{Graells-Garrido et~al\mbox{.}}{2017}]%
        {graells2017effect}
\bibfield{author}{\bibinfo{person}{Eduardo Graells-Garrido},
  \bibinfo{person}{Leo Ferres}, \bibinfo{person}{Diego Caro}, {and}
  \bibinfo{person}{Loreto Bravo}.} \bibinfo{year}{2017}\natexlab{}.
\newblock \showarticletitle{The effect of Pok{\'e}mon Go on the pulse of the
  city: a natural experiment}.
\newblock \bibinfo{journal}{\emph{EPJ Data Science}} \bibinfo{volume}{6},
  \bibinfo{number}{1} (\bibinfo{year}{2017}), \bibinfo{pages}{23}.
\newblock


\bibitem[\protect\citeauthoryear{Haklay}{Haklay}{2010}]%
        {haklay2010good}
\bibfield{author}{\bibinfo{person}{Mordechai Haklay}.}
  \bibinfo{year}{2010}\natexlab{}.
\newblock \showarticletitle{How Good is Volunteered Geographical Information? A
  Comparative Study of OpenStreetMap and Ordnance Survey Datasets}.
\newblock \bibinfo{journal}{\emph{Environment and Planning B: Planning and
  Design}} \bibinfo{volume}{37}, \bibinfo{number}{4} (\bibinfo{year}{2010}),
  \bibinfo{pages}{682--703}.
\newblock


\bibitem[\protect\citeauthoryear{Hunter}{Hunter}{2007}]%
        {hunter2007matplotlib}
\bibfield{author}{\bibinfo{person}{John~D Hunter}.}
  \bibinfo{year}{2007}\natexlab{}.
\newblock \showarticletitle{Matplotlib: A 2D Graphics Environment}.
\newblock \bibinfo{journal}{\emph{Computing in Science \& Engineering}}
  \bibinfo{volume}{9}, \bibinfo{number}{3} (\bibinfo{year}{2007}),
  \bibinfo{pages}{90}.
\newblock


\bibitem[\protect\citeauthoryear{i~LGTBI de l’Ajuntament~de
  Barcelona}{i~LGTBI de l’Ajuntament~de Barcelona}{2016}]%
        {i2016plan}
\bibfield{author}{\bibinfo{person}{Regidoria de~Feminismes i~LGTBI de
  l’Ajuntament~de Barcelona}.} \bibinfo{year}{2016}\natexlab{}.
\newblock \showarticletitle{Plan para la Justicia de G{\'e}nero 2016-2020}.
\newblock  (\bibinfo{year}{2016}).
\newblock
\urldef\tempurl%
\url{http://hdl.handle.net/11703/98743}
\showURL{%
\tempurl}


\bibitem[\protect\citeauthoryear{Marshall}{Marshall}{2004}]%
        {marshall2004transforming}
\bibfield{author}{\bibinfo{person}{Tim Marshall}.}
  \bibinfo{year}{2004}\natexlab{}.
\newblock \bibinfo{booktitle}{\emph{Transforming Barcelona: The Renewal of a
  European Metropolis}}.
\newblock \bibinfo{publisher}{Routledge}.
\newblock


\bibitem[\protect\citeauthoryear{McGuckin and Murakami}{McGuckin and
  Murakami}{1999}]%
        {mcguckin1999examining}
\bibfield{author}{\bibinfo{person}{Nancy McGuckin} {and}
  \bibinfo{person}{Elaine Murakami}.} \bibinfo{year}{1999}\natexlab{}.
\newblock \showarticletitle{Examining Trip-Chaining Behavior: Comparison of
  Travel by Men and Women}.
\newblock \bibinfo{journal}{\emph{Transportation Research Record}}
  \bibinfo{volume}{1693}, \bibinfo{number}{1} (\bibinfo{year}{1999}),
  \bibinfo{pages}{79--85}.
\newblock


\bibitem[\protect\citeauthoryear{McKinney et~al\mbox{.}}{McKinney
  et~al\mbox{.}}{2011}]%
        {mckinney2011pandas}
\bibfield{author}{\bibinfo{person}{Wes McKinney} {et~al\mbox{.}}}
  \bibinfo{year}{2011}\natexlab{}.
\newblock \showarticletitle{pandas: a Foundational Python Library for Data
  Analysis and Statistics}.
\newblock \bibinfo{journal}{\emph{Python for High Performance and Scientific
  Computing}} \bibinfo{volume}{14}, \bibinfo{number}{9} (\bibinfo{year}{2011}).
\newblock


\bibitem[\protect\citeauthoryear{McNeill, Bright, and Hale}{McNeill
  et~al\mbox{.}}{2017}]%
        {mcneill2017estimating}
\bibfield{author}{\bibinfo{person}{Graham McNeill}, \bibinfo{person}{Jonathan
  Bright}, {and} \bibinfo{person}{Scott~A Hale}.}
  \bibinfo{year}{2017}\natexlab{}.
\newblock \showarticletitle{Estimating local commuting patterns from geolocated
  Twitter data}.
\newblock \bibinfo{journal}{\emph{EPJ Data Science}} \bibinfo{volume}{6},
  \bibinfo{number}{1} (\bibinfo{year}{2017}), \bibinfo{pages}{24}.
\newblock


\bibitem[\protect\citeauthoryear{Metz}{Metz}{2000}]%
        {metz2000mobility}
\bibfield{author}{\bibinfo{person}{David~H Metz}.}
  \bibinfo{year}{2000}\natexlab{}.
\newblock \showarticletitle{Mobility of older people and their quality of
  life}.
\newblock \bibinfo{journal}{\emph{Transport Policy}} \bibinfo{volume}{7},
  \bibinfo{number}{2} (\bibinfo{year}{2000}), \bibinfo{pages}{149--152}.
\newblock


\bibitem[\protect\citeauthoryear{Milano, Novelli, and Cheer}{Milano
  et~al\mbox{.}}{2019}]%
        {milano2019overtourism}
\bibfield{author}{\bibinfo{person}{Claudio Milano}, \bibinfo{person}{Marina
  Novelli}, {and} \bibinfo{person}{Joseph~M Cheer}.}
  \bibinfo{year}{2019}\natexlab{}.
\newblock \showarticletitle{Overtourism and degrowth: a social movements
  perspective}.
\newblock \bibinfo{journal}{\emph{Journal of Sustainable Tourism}}
  \bibinfo{volume}{27}, \bibinfo{number}{12} (\bibinfo{year}{2019}),
  \bibinfo{pages}{1857--1875}.
\newblock


\bibitem[\protect\citeauthoryear{Monroe, Colaresi, and Quinn}{Monroe
  et~al\mbox{.}}{2008}]%
        {monroe2008fightin}
\bibfield{author}{\bibinfo{person}{Burt~L Monroe}, \bibinfo{person}{Michael~P
  Colaresi}, {and} \bibinfo{person}{Kevin~M Quinn}.}
  \bibinfo{year}{2008}\natexlab{}.
\newblock \showarticletitle{Fightin' Words: Lexical Feature Selection and
  Evaluation for Identifying the Content of Political Conflict}.
\newblock \bibinfo{journal}{\emph{Political Analysis}} \bibinfo{volume}{16},
  \bibinfo{number}{4} (\bibinfo{year}{2008}), \bibinfo{pages}{372--403}.
\newblock


\bibitem[\protect\citeauthoryear{Moran}{Moran}{1948}]%
        {moran1948interpretation}
\bibfield{author}{\bibinfo{person}{Patrick~AP Moran}.}
  \bibinfo{year}{1948}\natexlab{}.
\newblock \showarticletitle{The Interpretation of Statistical Maps}.
\newblock \bibinfo{journal}{\emph{Journal of the Royal Statistical Society.
  Series B (Methodological)}} \bibinfo{volume}{10}, \bibinfo{number}{2}
  (\bibinfo{year}{1948}), \bibinfo{pages}{243--251}.
\newblock


\bibitem[\protect\citeauthoryear{Noulas, Scellato, Lambiotte, Pontil, and
  Mascolo}{Noulas et~al\mbox{.}}{2012}]%
        {noulas2012tale}
\bibfield{author}{\bibinfo{person}{Anastasios Noulas},
  \bibinfo{person}{Salvatore Scellato}, \bibinfo{person}{Renaud Lambiotte},
  \bibinfo{person}{Massimiliano Pontil}, {and} \bibinfo{person}{Cecilia
  Mascolo}.} \bibinfo{year}{2012}\natexlab{}.
\newblock \showarticletitle{A Tale of Many Cities: Universal Patterns in Human
  Urban Mobility}.
\newblock \bibinfo{journal}{\emph{PLoS ONE}} \bibinfo{volume}{7},
  \bibinfo{number}{5} (\bibinfo{year}{2012}), \bibinfo{pages}{e37027}.
\newblock


\bibitem[\protect\citeauthoryear{of~the United Nations. Department~of Economic
  {and} Social Affairs (UN~DESA)}{of~the United Nations. Department~of Economic
  {and} Social Affairs (UN~DESA)}{2018}]%
        {UNpopulationdivision2018}
\bibfield{author}{\bibinfo{person}{Population~Division of~the United Nations.
  Department~of Economic {and} Social Affairs (UN~DESA)}.}
  \bibinfo{year}{2018}\natexlab{}.
\newblock \showarticletitle{The 2018 {Revision} of the {World} {Urbanization}
  {Prospects}}.
\newblock  (\bibinfo{year}{2018}).
\newblock


\bibitem[\protect\citeauthoryear{Perez}{Perez}{2019}]%
        {perez2019invisible}
\bibfield{author}{\bibinfo{person}{Caroline~Criado Perez}.}
  \bibinfo{year}{2019}\natexlab{}.
\newblock \bibinfo{booktitle}{\emph{Invisible Women: Exposing Data Bias in a
  World Designed for Men}}.
\newblock \bibinfo{publisher}{Random House}.
\newblock


\bibitem[\protect\citeauthoryear{Rey and Anselin}{Rey and Anselin}{2010}]%
        {rey2010pysal}
\bibfield{author}{\bibinfo{person}{Sergio~J. Rey} {and} \bibinfo{person}{Luc
  Anselin}.} \bibinfo{year}{2010}\natexlab{}.
\newblock \showarticletitle{PySAL: A Python Library of Spatial Analytical
  Methods}.
\newblock \bibinfo{journal}{\emph{Handbook of Applied Spatial Analysis}}
  (\bibinfo{year}{2010}), \bibinfo{pages}{175--193}.
\newblock


\bibitem[\protect\citeauthoryear{Ritchie and Roser}{Ritchie and Roser}{2018}]%
        {ritchie2018urbanization}
\bibfield{author}{\bibinfo{person}{Hannah Ritchie} {and} \bibinfo{person}{Max
  Roser}.} \bibinfo{year}{2018}\natexlab{}.
\newblock \showarticletitle{Urbanization}.
\newblock \bibinfo{journal}{\emph{Our World in Data}} (\bibinfo{year}{2018}).
\newblock


\bibitem[\protect\citeauthoryear{Vasquez-Henriquez, Graells-Garrido, and
  Caro}{Vasquez-Henriquez et~al\mbox{.}}{2019}]%
        {vasquez2019characterizing}
\bibfield{author}{\bibinfo{person}{Paula Vasquez-Henriquez},
  \bibinfo{person}{Eduardo Graells-Garrido}, {and} \bibinfo{person}{Diego
  Caro}.} \bibinfo{year}{2019}\natexlab{}.
\newblock \showarticletitle{Characterizing Transport Perception using Social
  Media: Differences in Mode and Gender}. In
  \bibinfo{booktitle}{\emph{Proceedings of the 10th ACM Conference on Web
  Science}}. \bibinfo{pages}{295--299}.
\newblock


\bibitem[\protect\citeauthoryear{Ziegler and Schwanen}{Ziegler and
  Schwanen}{2011}]%
        {ziegler2011like}
\bibfield{author}{\bibinfo{person}{Friederike Ziegler} {and}
  \bibinfo{person}{Tim Schwanen}.} \bibinfo{year}{2011}\natexlab{}.
\newblock \showarticletitle{‘I like to go out to be energised by different
  people’: an exploratory analysis of mobility and wellbeing in later life}.
\newblock \bibinfo{journal}{\emph{Ageing \& Society}} \bibinfo{volume}{31},
  \bibinfo{number}{5} (\bibinfo{year}{2011}), \bibinfo{pages}{758--781}.
\newblock


\end{thebibliography}

\end{document}